# Plackett-Luce regression:
# A new Bayesian model for polychotomous data

**Cédric Archambeau**
Xerox Research Center Europe
Meylan, France
cedric.archambeau@xrce.xerox.com

**François Caron**
INRIA
Univ. Bordeaux, IMB, UMR 5251
Talence, France
francois.caron@inria.fr

## Abstract

Multinomial logistic regression is one of the most popular models for modelling the effect of explanatory variables on a subject choice between a set of specified options. This model has found numerous applications in machine learning, psychology or economy. Bayesian inference in this model is non trivial and requires, either to resort to a Metropolis-Hastings algorithm, or rejection sampling within a Gibbs sampler. In this paper, we propose an alternative model to multinomial logistic regression. The model builds on the Plackett-Luce model, a popular model for multiple comparisons. We show that the introduction of a suitable set of auxiliary variables leads to an Expectation-Maximization algorithm to find Maximum A Posteriori estimates of the parameters. We further provide a full Bayesian treatment by deriving a Gibbs sampler, which only requires to sample from highly standard distributions. We also propose a variational approximate inference scheme. All are very simple to implement. One property of our Plackett-Luce regression model is that it learns a sparse set of feature weights. We compare our method to sparse Bayesian multinomial logistic regression and show that it is competitive, especially in presence of polychotomous data.

## 1 Introduction

The multinomial logistic regression (or multinomial logit) model is one of the most popular models for modelling the effect of explanatory variables $X_i = (X_{i1}, \ldots, X_{id}) \in \mathcal{X}$ on a subject choice $Y_i$ between a set of prespecified options $\{1, \ldots, K\}$. The model, which belongs to the class of generalized linear models (McCullagh & Nelder, 1989), takes the following form:

$$\Pr(Y_i = k | X_i, \beta) = \frac{e^{X_i^\top \beta_k}}{1 + \sum_{\ell=1}^{K-1} e^{X_i^\top \beta_\ell}} \qquad (1)$$

for $k = 1, \ldots, K-1$ and $\Pr(Y_i = K | X_i) = \frac{1}{1+\sum_{\ell=1}^{K-1} \exp(X_i^\top \beta_\ell)}$. The monograph of Agresti (1990) gives details on the foundations of this model and related ones such as conditional logit (McFadden, 1973). The parameters $\beta_k = (\beta_{k1}, \ldots, \beta_{kd})$, $k = 1, \ldots, K-1$ are unknown and have to be estimated from the data. Bayesian inference in multinomial logit is complicated by the fact that no conjugate prior exists for the regression parameters. Various algorithms have been proposed in the literature, see e.g. (Dey et al., 2000). In particular, similarly to Bayesian inference in probit models (Albert & Chib, 1993; Talhouk et al., 2012; Holmes & Held, 2006) proposed an auxiliary variable model for block Gibbs sampling. Alternative Markov Chain Monte Carlo (MCMC) algorithms have been compared by Girolami & Calderhead (2011). However, although some algorithms show excellent mixing properties, they lack simplicity of implementation.

When only some of the predictors are assumed to be relevant, sparse multinomial logistic regression has been proposed using Laplace priors.Krishnapuram et al. (2005) obtained Maximum A Posteriori (MAP) estimates via Minorization-Maximization-based algorithms, while fully Bayesian inference is conducted in (Cawley et al., 2007) and (Genkin et al., 2007).

In this article, we propose an alternative model to multinomial logit for multi-class classification and discrete choice modelling. We show that the use of a carefully chosen set of latent variables leads to an Expectation Maximization (EM) algorithm to find MAP estimates, a Gibbs sampler or a variational EM algorithm to approximate the full posterior. Importantly, the Gibbs sampler only requires to sample from standard distributions, which makes it computationally highly amenable. Moreover, for some values of the hyperpa-rameter, we show that the model induces sparsity over the parameters. In particular, the MAP estimates are exactly sparse. We also provide detailed comparaisons with Bayesian sparse multinomial logit in terms of computational efficiency and prediction performances. The Bayesian treatment of sparse multinomial logistic regression is discussed in Appendix A. The model builds on the data augmentation model proposed by Holmes & Held (2006) and the Bayesian lasso of Park & Casella (2008). To the best of our knowledge is has not been proposed before.

## 2 Statistical model

We propose an alternative model for categorical data analysis. The model is defined as follows:

$$\Pr(Y_i = k | X_i, \lambda) = \frac{W_i^\top \lambda_k}{\sum_{\ell=1}^K W_i^\top \lambda_\ell} \quad (2)$$

for $k = 1, \ldots, K$, where $W_i = (W_{i1}, \ldots, W_{ip})$ and $W_{ij} = \mathcal{K}_j(X)$. $\mathcal{K}_j$ is a known function from $\mathcal{X}$ to $[0, +\infty[$. In practice, there is a lot of flexibility on the choice of the transformations $\mathcal{K}_j$ of the explanatory variables. In the remaining of this paper, we will consider $\mathcal{K}_j(X) = \exp(X_{ij})$, $\mathcal{K}_{d+j}(X) = \exp(-X_{ij})$ to account for negative effects and $\mathcal{K}_{2d+1}(X) = \exp(0)$ for the offset. The non-negative parameters $\lambda_{kj}$, $k = 1, \ldots, K$, $j = 1, \ldots, p$ are unknown and have to be estimated from the data.

The model shares strong similarities with the Plackett-Luce model for multiple comparisons (Luce, 1959; Plackett, 1975), for which efficient Bayesian inference can be carried out (Guiver & Snelson, 2009; Caron & Doucet, 2012). Similarly to this model, it also admits a Thurstonian interpretation (Diaconis, 1988) that we describe now. For $k = 1, \ldots, K$ and $j = 1, \ldots, p$, let

$$V_{ikj} \sim \text{Exp}(W_{ij} \lambda_{kj}) ,$$

where $\text{Exp}(b)$ denotes the exponential distribution with rate parameter $b$. As an analogy, let consider a race between $K$ different teams, each team having $p$ competitors. Each individual $j$ in a team $k$ has an arrival time $V_{ikj}$ for competition $i$. Then $Y_i$ denotes the winning team, i.e. the team $k$ which has the individual with the lowest arrival time. Following properties of the exponential distribution, it is straightforward to show that

$$\Pr(\arg\min_\kappa(\min_j V_{i\kappa j}) = k | X_i, \lambda) = \frac{W_i^\top \lambda_k}{\sum_{\ell=1}^K W_i^\top \lambda_\ell} .$$

The decision boundaries between two classes $k$ and $\ell$ are given by
$$W^\top(\lambda_k - \lambda_\ell) = 0 ,$$

i.e. they are linear in the transformed domain $W$. As the multinomial logit model, the model (2) satisfies Luce's axiom of choice (Luce, 1977) a.k.a. *independence from irrelevant alternatives*.

## 3 MAP estimation and Bayesian Inference

### 3.1 Data Augmentation

The likelihood defined by (2) does not admit a conjugate prior. We can nonetheless consider a data augmentation scheme as indicated by Caron & Doucet (2012). Assume that we have observed data $\mathcal{Y} = \{Y_i\}_{i=1}^n$. We introduce auxiliary latent variables $\mathcal{C} = \{C_i\}_{i=1}^n$ and $\mathcal{Z} = \{Z_i\}_{i=1}^n$ such that, for $i = 1, \ldots, n$

$$Y_i | C_i, \lambda \sim \text{Disc}\left(\frac{\lambda_{1C_i}}{\sum_\ell \lambda_{\ell C_i}}, \ldots, \frac{\lambda_{KC_i}}{\sum_\ell \lambda_{\ell C_i}}\right), \quad (3)$$

$$C_i | \lambda \sim \text{Disc}\left(\frac{W_{i1} \sum_k \lambda_{k1}}{W_i^\top \sum_\ell \lambda_\ell}, \ldots, \frac{W_{ip} \sum_k \lambda_{kp}}{W_i^\top \sum_\ell \lambda_\ell}\right), \quad (4)$$

$$Z_i | \lambda \sim \text{Exp}\left(W_i^\top \sum_\ell \lambda_\ell\right), \quad (5)$$

where $\text{Disc}(\pi_1, \ldots, \pi_p)$ denotes the discrete distribution of parameters $(\pi_1, \ldots, \pi_p)$ where $\pi_i \geq 0$ and $\sum_i \pi_i = 1$. Pursuing the analogy with team competition introduced in the previous section, the latent variables $Z_i$ and $C_i$ have the following interpretation. $Z_i = \min_{k,j}(V_{ijk})$ corresponds to the arrival time of the winner of the competition. As the variables $V_{ijk}$ are exponentially distributed and the minimum of two exponential variable of rates $w_1$ and $w_2$ is an exponential variables of rate $w_1 + w_2$, we recover (5). $C_i \in \{1, \ldots, p\}$ corresponds to the index of the individual in the winning team who arrives first.

The model (3–4) corresponds to an ad-mixture of Discrete distributions. Indeed, the class probabilities defined by the Plackett-Luce model (2) can be rewritten in the form of a mixture model by integrating out the latent indicator variables $\{C_i\}_{i=1}^n$:

$$\Pr(Y_i = k | X_i, \lambda) = \sum_j \pi_{ij} \text{Disc}\left(\frac{\lambda_{1j}}{\sum_\ell \lambda_{\ell j}}, \ldots, \frac{\lambda_{Kj}}{\sum_\ell \lambda_{\ell j}}\right),$$

where the mixture weight $\pi_{ij} = \frac{W_{ij} \sum_k \lambda_{kj}}{W_i^\top \sum_\ell \lambda_\ell}$ depends on the data through $\{W_{ij}\}_{j=1}^p$. Note that there is one mixture component per feature. Hence, each component characterises the classes by assigning a different importance weight to their associated feature.

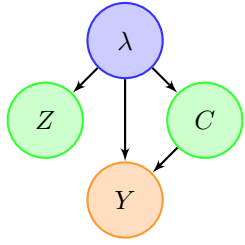

Figure 1: Graphical Model. Parameter of interest $\lambda$ is in blue, latent variables $Z$ and $C$ in green and measurements $Y$ in orange.

The resulting log-complete likelihood is given by

$$\ln p(\mathcal{Y}, \mathcal{C}, \mathcal{Z}|\lambda) = \sum_{k=1}^{K}\sum_{j=1}^{p} \left\{ n_{kj} \ln \lambda_{kj} - \lambda_{kj} \sum_{i=1}^{n} Z_i W_{ij} \right\}$$
$$+ \sum_{i=1}^{n}\sum_{j=1}^{p} \delta_{C_i j} \ln W_{ij} \ , \qquad (6)$$

where $n_{kj} = \sum_i \delta_{Y_i k} \delta_{C_i j}$ and $\delta_{ij}$ is the Kronecker delta. We further assign a conjugate Gamma prior to the parameters $\lambda$:

$$\lambda \sim \prod_{k=1}^{K}\prod_{j=1}^{p} \mathrm{Gam}(\lambda_{kj}; a, b) \ . \qquad (7)$$

where $a > 0$ is the shape parameter and $b > 0$ is the rate parameter. The graphical model is shown in Figure 1.

### 3.2 EM algorithm

The log-posterior can be maximized using the EM algorithm by iteratively maximizing a lower bound, called the negative *variational free energy* (Neal & Hinton, 1998), or a penalized version as below:

$$\ln p(\lambda|\mathcal{Y}) \geq \langle \ln p(\mathcal{Y}, \mathcal{C}, \mathcal{Z}|\lambda) \rangle_q + \mathrm{H}\left(q(\mathcal{C}, \mathcal{Z})\right) + \ln p(\lambda)$$
$$\equiv -\mathcal{F}(q, \lambda) + \ln p(\lambda) \ , \qquad (8)$$

where $\langle \cdot \rangle_q$ denotes the expectation with respect to $q(\mathcal{C}, \mathcal{Z})$ and $\mathrm{H}(\cdot)$ is the (differential) entropy. When the posterior $p(\mathcal{C}, \mathcal{Z}|\mathcal{Y}, \lambda)$ is analytically tractable, then we set $q(\mathcal{C}, \mathcal{Z}) = p(\mathcal{C}, \mathcal{Z}|\mathcal{Y}, \lambda)$ so that the bound is exact and we recover the EM algorithm.

Given the augmented model (3-5), the posterior factorizes: $p(\mathcal{C}, \mathcal{Z}|\mathcal{Y}, \lambda) = \prod_i p(C_i|Y_i, \lambda) p(Z_i|\mathcal{Y}, \lambda)$. The E-step consists in computing the posterior of the latent indicator variables for fixed $\lambda$.[1] This leads to

$$C_i | Y_i, \lambda \sim \mathrm{Disc}\left( \frac{W_{i1} \lambda_{Y_i 1}}{W_i^\top \lambda_{Y_i}}, \ldots, \frac{W_{ip} \lambda_{Y_i p}}{W_i^\top \lambda_{Y_i}} \right) . \quad (9)$$

---
[1]Note that $p(Z_i|\mathcal{Y}, \lambda)$ is given by (5)

The M-step updates the parameter set $\lambda$ for fixed posteriors:

$$\lambda \leftarrow \arg\max_{\lambda} \langle \ln p(\mathcal{Y}, \mathcal{C}|\lambda) \rangle_{p(\mathcal{C}|\mathcal{Y}, \lambda)} + \ln p(\lambda) \ .$$

It follows that

$$\lambda_{kj} = \begin{cases} \frac{a - 1 + \langle n_{kj} \rangle}{b + \sum_i \langle z_i \rangle W_{ij}} & \text{if } a > 1 - \langle n_{kj} \rangle , \\ 0 & \text{otherwise.} \end{cases} \quad (10)$$

where $\langle n_{kj} \rangle = \sum_i \frac{W_{ij} \lambda_{kj}}{W_i^\top \lambda_k}$ and $\langle z_i \rangle = \left( W_i^\top \sum_l \lambda_l \right)^{-1}$. For $a = 1$ and $b = 0$, the maximum a posteriori estimate coincides with the maximum likelihood estimate. If $a < 1$, one may obtain sparse estimates of the weights, as the numerator of (10) may become negative. By changing the value of the hyperparameter $a$, one may obtain different levels of sparsity. It could for example be set by cross-validation. Next, we propose a Bayesian treatment of $\lambda$. We first show that it is straightforward to derive a Gibbs sampler for (2) given our data augmentation. Subsequently, we derive a deterministic approximate Bayesian inference scheme, closely related to the Gibbs sampler.

### 3.3 Gibbs sampler

Using the same data augmentation, we can define a Gibbs sampler for sampling from the posterior distribution $p(\mathcal{C}, \mathcal{Z}, \lambda|\mathcal{Y})$. The conditional distribution $p(\mathcal{C}|\mathcal{Y}, \lambda)$ factorizes, such that each $C_i$ can be updated conditioned on $(Y_i, \lambda)$ using (9). Likewise, the conditional $p(\mathcal{Z}|\mathcal{Y}, \lambda)$ factorizes, such that each $Z_i$ can be updated conditioned on $\lambda$ as the posterior reverts to the prior (5). Finally, the conditional for each $\lambda_{kj}$ is of the form $p(\lambda_{kj}|\mathcal{Y}, \mathcal{C}, \mathcal{Z}) \propto p(\lambda_{kj}) e^{\ln p(\mathcal{Y}, \mathcal{C}, \mathcal{Z}|\lambda)}$. Since the Gamma prior is conjugate to the complete likelihood, the conditional for $\lambda_{kj}$ still follows a Gamma distribution.

To summarize, the Gibbs sampler is given by

$$C_i|Y_i, \lambda \sim \mathrm{Disc}\left( \frac{W_{i1} \lambda_{Y_i 1}}{W_i^\top \lambda_{Y_i}}, \ldots, \frac{W_{ip} \lambda_{Y_i p}}{W_i^\top \lambda_{Y_i}} \right), \quad (11)$$

$$Z_i|\lambda \sim \mathrm{Exp}\left( W_i^\top \sum_{\ell=1}^{K} \lambda_\ell \right), \quad (12)$$

$$\lambda_{kj}|\mathcal{Y}, \mathcal{C}, \mathcal{Z} \sim \mathrm{Gam}\left( a + n_{kj}, b + \sum_{i=1}^{n} Z_i W_{ij} \right). \quad (13)$$

### 3.4 Variational approximation

Next, we turn our attention to a deterministic approximation instead of sampling. Variational EM maximizes the variational lower bound (Neal & Hinton,

1998), which can be re-written in the form

$$\ln p(\mathcal{Y}) \geq -\mathcal{F}(q) = \langle \ln p(\mathcal{Y}, \mathcal{C}, \mathcal{Z}, \lambda) \rangle_{q(\mathcal{C})q(\mathcal{Z})q(\lambda)} \quad (14)$$
$$+ \text{KL}\left[q(\mathcal{C})q(\mathcal{Z})q(\lambda)\|p(\mathcal{C}, \mathcal{Z}, \lambda|\mathcal{Y}))\right].$$

The approximation assumes that the latent variables and the parameters are independent a posteriori given the data. The approximate posteriors are given by

$$q(\mathcal{C}) \propto p(\mathcal{C})e^{\langle \ln p(\mathcal{Y}, \mathcal{Z}, \lambda|\mathcal{C}) \rangle_{q(\mathcal{Z})q(\lambda)}}, \quad (15)$$
$$q(\mathcal{Z}) \propto e^{\langle \ln p(\mathcal{Y}, \mathcal{C}, \mathcal{Z}, \lambda) \rangle_{q(\mathcal{C})q(\lambda)}}, \quad (16)$$
$$q(\lambda) \propto p(\lambda)e^{\langle \ln p(\mathcal{Y}, \mathcal{C}, \mathcal{Z}|\lambda) \rangle_{q(\mathcal{C})q(\mathcal{Z})}}, \quad (17)$$

In practice, this boils down to cycling through the following updates

$$\rho_{kji} \propto \delta_{Y_i k} W_{ij} e^{\langle \ln \lambda_{kj} \rangle}, \quad \langle z_i \rangle = \frac{1}{W_i^\top \sum_\ell \langle \lambda_\ell \rangle}, \quad (18)$$
$$a_{kj} = a + \langle n_{kj} \rangle, \qquad b_{kj} = b + \sum_i \langle z_i \rangle W_{ij}, (19)$$

where $\langle \lambda_{kj} \rangle = \frac{a_{kj}}{b_{kj}}$, $\langle \ln \lambda_{kj} \rangle = \psi(a_{kj}) - \ln b_{kj}$, and $\langle n_{kj} \rangle = \sum_i \rho_{kji}$.

The similarity between the variational posteriors and the Gibbs sampler is striking. For example, $p(\lambda|\mathcal{Y}, \mathcal{C}, \mathcal{Z})$ and $q(\lambda)$ have the same form, except that the counts $n_{kj}$ and the latent auxiliary variables $Z_i$ are respectively replaced by their expected values, namely $\langle n_{kj} \rangle$ and $\langle Z_i \rangle$. However, unlike Gibbs sampling, the convergence of variational inference and the correctness of the algorithm are easy to check by monitoring the variational lower bound, which monotonically increases at each iteration. Hence, considering both approaches in parallel is convenient when debugging the Gibbs sampling code.

### 3.5 Identifiability and hyperparameters estimation

The likelihood (2) is invariant to a rescaling of the $\lambda_{kj}$'s. As a consequence, the scaling hyperparameter $b$ does not have any effect on the prediction, It can be set to an arbitrary value, e.g. $b = 1$. Let

$$\Lambda = \sum_k \sum_j \lambda_{kj}, \qquad \overline{\lambda_{kj}} = \frac{\lambda_{kj}}{\Lambda}.$$

Because of the invariance w.r.t. rescaling, $\Lambda$ is not likelihood identifiable and

$$p(\Lambda, \overline{\lambda}|\mathcal{Y}) = p(\Lambda)p(\overline{\lambda}|\mathcal{Y}),$$

with $p(\Lambda) = \text{Gam}(\Lambda; Kp, b)$. It follows that

$$\Lambda^{\text{MAP}} = \frac{Kp - 1}{b}.$$

To improve the mixing of the Markov chain, an additional sampling step can be added by sampling the total mass $\Lambda$ from the prior then rescale the parameters $\lambda$ adequately. While it would improve the mixing of the Markov chain, this is useless here as we are typically interested in the prediction with the normalized weights.

The parameter $a$ can be estimated by cross-validation in the EM algorithm. For full Bayesian inference, we assume the following flat improper prior:

$$p(a) \propto \frac{1}{a}$$

and we add a Metropolis-Hastings sampling step in the Gibbs sampler. In the variational EM algorithm, $a$ is estimated by type II Maximum Likelihood:

$$a \leftarrow \arg\max_a \left\{ -\mathcal{F}(q, \lambda) + \ln p(a) \right\}.$$

While this does not lead to a closed form update for $a$, the maximization can be performed by a line search.

## 4 Experiments

In this section we investigate the performances of the Plackett-Luce regression model, which we will denote PL-EM, PL-Gibbs or PL-Var depending on the training algorithm we used (see Section 3).

### 4.1 Sparsity and regularization paths

First, we compare the sparsity properties of PL-EM, PL-Gibbs and PL-Var. We consider the iris dataset, which is available from the UCI repository. As with $\ell_1$-penalization in generalized linear regression models, varying the value of the regularization coefficient $a$ will enforce the amount of sparsity over the estimated $\lambda$ of our model.

Figure 2 shows the regularization path for the coefficients as a function of $a$, using the MAP estimate obtained with EM, the posterior mean and median obtained with Gibbs sampling, and the posterior mean obtained with the variational approximation. Regularization paths report the values of the weights obtained when training the models with decreasing values of the hyperparameter $a$. When considering MAP, we obtain exactly sparse estimates for sufficiently small values of $a$. Similarly to what is observed with the Bayesian lasso (Park & Casella, 2008) the posterior mean estimates obtained by Gibbs sampling are not sparse but more concentrated around zero as the value of $a$ decreases. The posterior median leads to sparser estimates, similarly to what is observed with the Bayesian lasso. Interestingly, the variational approximation shows similar sparsity as the posterior median, but converges faster.

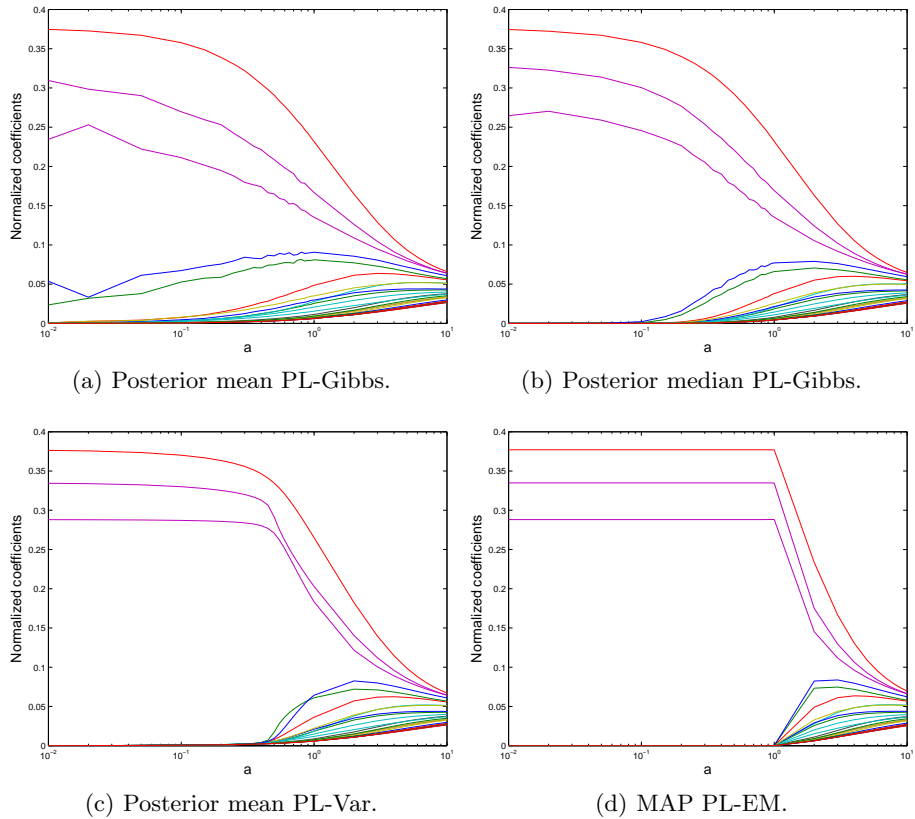

(a) Posterior mean PL-Gibbs.

(b) Posterior median PL-Gibbs.

(c) Posterior mean PL-Var.

(d) MAP PL-EM.

Figure 2: Regularization paths for the iris dataset. The paths are the values of the weight parameters as a function of $a$. They are obtained by training the models for decreasing values of $a$. Each colour corresponds to a different $\lambda_{kj}$. The coefficient values are normalized to make the scales comparable.

### 4.2 Toy datasets

Next, we investigate the predictive and computational performances of the proposed PL models. We compare to Bayesian multinomial logistic regression (or logit). More specifically, we consider the Gibbs sampler described by Holmes & Held (2006), as the authors provide detailed pseudo-code on the algorithm, and the algorithm has no tuning parameter as in the case of our Gibbs sampler (as opposed to hybrid Monte Carlo, for example).[2] However, we will consider a sparse extension inspired by the Bayesian lasso of Park & Casella (2008). The Gibbs sampler is detailed in the Appendix.

We compare both models/algorithms in terms of computational efficiency and predictive performance on five datasets from the UCI repository: three polychotomous (Lenses, Wine and Iris) and three binary (Heart, German and Pima). Summaries of the different datasets are given in Table 1.

| Name | $d$ | $K$ | $n$ |
|---|---|---|---|
| Lenses | 4 | 3 | 24 |
| Wine | 13 | 3 | 178 |
| Iris | 4 | 3 | 150 |
| Heart | 13 | 2 | 270 |
| German | 24 | 2 | 1000 |
| Pima | 8 | 2 | 768 |

Table 1: Dataset characteristics: covariates ($d$), categories ($K$), and data points ($n$).

As Holmes & Held (2006), we first conduct 5000 burn in iterations then the next 5000 iterations are used to collect posterior samples. Each method was implemented in Matlab on a standard computer. We compare the predictive performance by computing the average number of misclassifications over 20 replications (i.e. random splits of the data in training and test set). Predictions are based on Bayesian averaging over 5000 samples. The results are reported in Table 2. It can be observed that the proposed Plackett-Luce models are competitive with sparse Bayesian multinomial logit.

---

[2]To be precise we used the corrected version by van der Lans (2011); Holmes & Held (2011).

This is confirmed when computing the area under the receiver operating curve (Figure 3). Gibbs sampling is in general beneficial compared to the variational approximation. This can be explained by the fact the PL-Var tends to provide sparser solutions, hence loosing predictive power.

We used the same experimental setup to investigate the relative efficiency of PL-Gibbs and sparse Bayesian multinomial logit. The two methods are compared by calculating the effective sample size using the posterior samples for each covariate

$$ESS = \frac{N}{1 + 2\sum_k \gamma(k)} \ ,$$

where $N = 5000$ and $\sum_k \gamma(k)$ is the sum of the $K$ monotone sample auto-correlations as estimated by the initial monotone sequence estimator(Geyer, 1992; Girolami & Calderhead, 2011). We report the minimum ESS over the set of whole set of covariates. Table 3 shows the results for the different datasets, based on 20 replications. It can be observed that ESS is better (higher) for sparse multinomial logit in the binary cases. However, the running time is also higher, leading to a relative speed (ratio of time to ESS) of 2 to 3 in favour of sparse logit. However, when the number of classes increases this trend is drastically changed with a relative speed of approximately 30 in favour of PL-Gibbs. As discussed in Section 5, we attribute this to the fact that, in multinomial logit, the coefficients of one class are sampled conditioned on the coefficients of the other classes. Unfortunately, these coefficients are strongly correlated, resulting in a poor mixing. By contrast, the PL regression models jointly sample (or update) the coefficients associated to all the classes.

### 4.3 Real data examples

First, we consider the colon cancer data[3]. This is a binary classification problem consisting of 62 data points and 2001 features. In general, 50 data points are used for training and 12 for testing. The average number of misclassifications is $0.36 \pm 0.11$ for PL-Gibbs and $0.33 \pm 0.15$ for sparse binary logit. Hence, both model perform similarly. However, sparse binary logit based on the sampler of Holmes & Held (2006) runs about 500 times slower.

Second, we consider the sushi data (Kamishima, 2003). Individuals were asked which out of 10 sushis they preferred. The number of training data points is 4000 and the number of test data points is 1000. The number of features is 11. We repeated the experiment 5 times. Here, the average number of misclassifications for Plackett-Luce and sparse multinomial logit were respectively 656 and 672. While the performances are poor in both cases, PL-Gibbs performs slightly better and is much faster to run. Generating 5000 samples with PL-Gibbs takes a couple of minutes on a standard quad-core laptop, while it takes approximately 6 hours with the sampler of Holmes & Held (2006).

---
[3] http://perso.telecom-paristech.fr/~gfort/GLM/Programs.html

| Dataset | Sparse Logit | PL-Gibbs | PL-Var |
|---|---|---|---|
| Pima | .240 (.024) | **.238** (.015) | .239 (.017) |
| Iris | **.086** (0.086) | .186 (.055) | .181 (.057) |
| Heart | .215 (.046) | **.170** (.023) | .223 (.056) |
| German | .262 (.021) | **.260** (.017) | .298 (.015) |
| Lenses | **.700** (0.087) | .825 (.071) | .821 (.073) |
| Wine | .080 (0.038) | **.048** (.019) | .093 (.048) |

Table 2: Average number of misclassifications (and standard deviations) for the different dataset.

| Dataset | Method | Time | ESS | $\frac{\text{Time}}{\text{ESS}}$ | Relat. Speed |
|---|---|---|---|---|---|
| Lenses | Sp. Logit | 26.0 | 304 | 0.086 | 1 |
|  | PL-Gibbs | 8.6 | 2916 | 0.003 | 29 |
| Wine | Sp. Logit | 184.9 | 8 | 22.960 | 1 |
|  | PL-Gibbs | 15.1 | 23 | 0.655 | 35 |
| Iris | Sp. Logit | 139.9 | 8 | 17.996 | 1 |
|  | PL-Gibbs | 10.8 | 14 | 0.747 | 24 |
| Heart | Sp. Logit | 128.8 | 270 | 0.477 | 2 |
|  | PL-Gibbs | 16.0 | 14 | 1.185 | 1 |
| German | Sp. Logit | 561.1 | 418 | 1.342 | 3 |
|  | PL-Gibbs | 58.0 | 17 | 3.426 | 1 |
| Pima | Sp. Logit | 409.6 | 675 | 0.607 | 2 |
|  | PL-Gibbs | 23.2 | 23 | 1.028 | 1 |

Table 3: Efficiency of Sparse multinomial logit and Plackett-Luce regression on the different datasets.

## 5 Discussion and extensions

The Gibbs sampler of the Plackett-Luce regression model only requires to sample from highly standard distributions (Exponential, Gamma and Discrete) for which very efficient generators exist. This is to be compared to Bayesian (multinomial) logistic regression, where it is required to sample from the truncated logistic and from other distributions without standard form, like for example the Kolmogorov-Smirnov, with rejection sampling (Holmes & Held, 2006). An important drawback of recent treatments of multinomial logistic regression like the one of Holmes & Held (2006) is that the multinomial outcomes are expressed as a sequence of binary outcomes. In other words, each vector of coefficient $\beta_k$ is sampled conditional on the others. This explains why we observed experimentally the relatively poor mixing properties of multinomial logit when the number of classes is more than 2. Inference in multinomial probit models often suffer from the same weakness (e.g., Albert & Chib, 1993). In the Gibbs sampler associated to the model we propose, the parameters $\lambda_{kj}$ are updated jointly and independently given the latent variables.

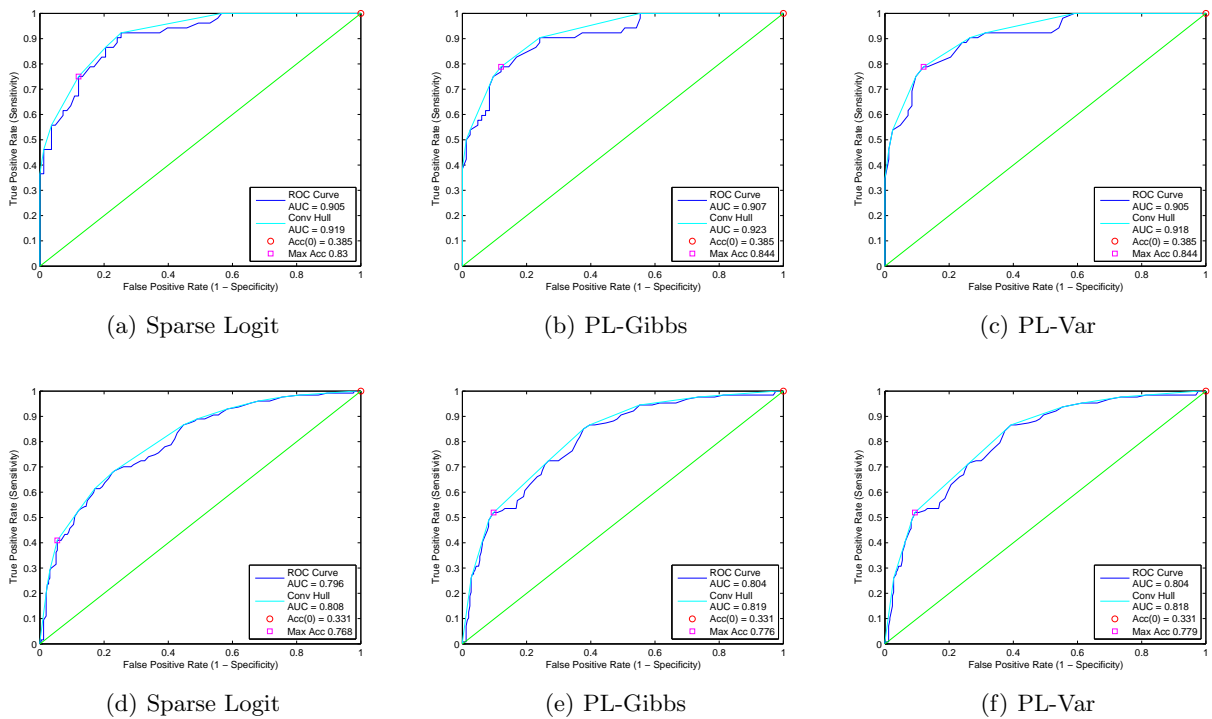

Figure 3: ROC for heart (first row) and pima (second row) datasets.

Throughout the paper, we arbitrarily used an exponential transformation of the covariates. Additional flexibility could be introduced into the model by allowing other transformations, which would better characterize the classes. Indeed, one of the limitations of the proposed generative model is that for each observation $X_i$, it is assumed first a feature $C_i$ is selected, and then its class $Y_i$ is drawn given $C_i$. When features do not uniquely characterize the classes, it becomes difficult to distinguish them. Figure 4 shows the decision boundaries for the iris data set (when only considering the petal length and width). It could be argued that these decision boundaries are counter-intuitive, especially in contrast to the decision boundaries of the sparse multinomial logit. This is confirmed by the poor performance of PL regression on the iris data set (see Table 2). However, by introducing additional transformation, more natural boundaries can be obtained. For example, to obtain Figure 4) we considered the additional features $\exp(X_{i1} + X_{i2})$ and $\exp(-X_{i1} - X_{i2})$.

In general, devicing features of this kind would require prior knowledge of the problem at hand, which is rarely the case.

In this paper, we have focused on a Gamma prior for the weights parameters $\lambda_{kj}$. Now assume that $a = \alpha/p$. When the number of covariates $p$ and hence $p^\star$ becomes very large, we have

$$\Lambda = \sum_{j=1}^{\infty} \lambda_{kj} \sim \text{Gam}(\alpha, b) \ .$$

Hence, the sequence of normalized ordered weights $\frac{\lambda_{k\sigma(1)}}{\Lambda} > \frac{\lambda_{k\sigma(2)}}{\Lambda} > \ldots$ is drawn from the Poisson Dirichlet distribution of parameter $\alpha$(Pitman, 1996), which is the distribution of the ordered weights in a draw from a Dirichlet process (Teh, 2010). The limit may give useful hints on the behavior of the model with a large number of covariates.

The model could also be directly extended by replacing the Gamma prior by the larger family of generalized inverse Gaussian distributions, which admits the Gamma distribution as a special case, and is also a conjugate prior for the complete-likelihood. Hence, the full conditional distribution of $\lambda_{kj}$ follows a generalized inverse Gaussian distribution. The use of this distribution may offer more flexibility in the modelling of the tails of the prior for the parameters $\lambda_{kj}$.

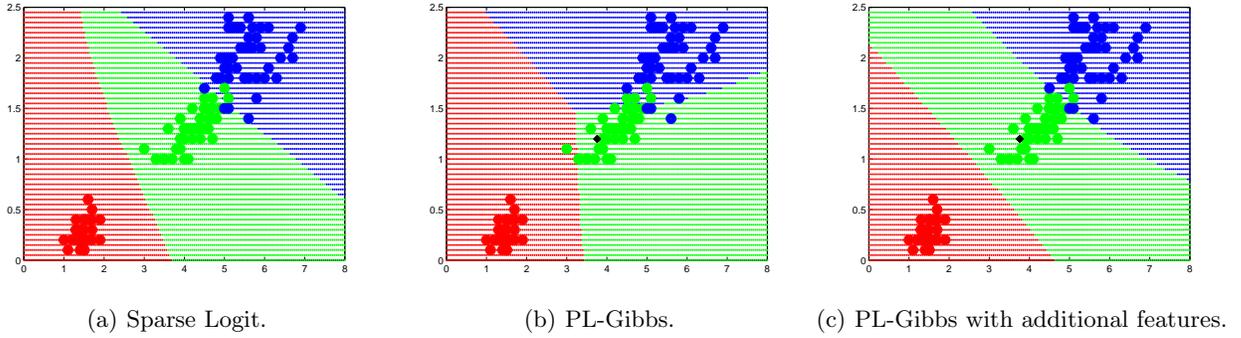

| (a) Sparse Logit. | (b) PL-Gibbs. | (c) PL-Gibbs with additional features. |

Figure 4: Decision boundaries for iris dataset (feature 3 and 4).

Finally, it should be noted that the model studied in this paper has been suggested by Caron & Doucet (2012). However, no discussion of the algorithms, nor any study on the fit of this model to data were discussed there.

## A   Bayesian sparse multinomial logistic regression

In this Appendix, we provide a Gibbs sampler for sparse multinomial logistic regression. The likelihood is given by (1) and a sparsity-promoting prior is imposed on the regression weights $\beta_{kj}$, $k = 1, \ldots, K - 1$ and $j = 1, \ldots, p$. More specifically, we consider the following hierarchical prior:

$$\beta_{kj}|\tau_{kj} \sim \mathcal{N}(0, \tau_{kj}), \qquad \tau_{kj} \sim \mathrm{Exp}\left(\frac{\theta}{2}\right).$$

This model ensures that marginally $\beta_{kj}$ follows a double-exponential (a.k.a. Laplace) distribution of parameter $\theta$. It is routinely used in Bayesian generalized linear mode to induce sparsity Genkin et al. (2007); Park & Casella (2008); Griffin & Brown (2010). We further assume a Gamma prior for the hyperparameter $\theta$ for conjugacy reasons (see below (20)):

$$\theta \sim \mathrm{Gam}(c, d) \ .$$

As noted by Park & Casella (2008), the improper prior $p(\theta) \propto \frac{1}{\theta}$ should be avoided in this model as it leads to an improper posterior for $\theta$.

We can define a Gibbs sampler for the above model by using auxiliary variables $u_k$, $k = 1, \ldots, K - 1$ as defined in (Holmes & Held, 2006). We therefore can define a Gibbs sampler to sample from the full posterior distribution $p(\beta, u, \tau, \theta|y)$. The sampler is defined as follows at each iteration:

- For $k = 1, \ldots, K - 1$:
  - Sample $u_k, \beta_k|\beta_{-k}, u_{-k}, \{\tau_{kj}\}$ as described in (Holmes & Held, 2006; van der Lans, 2011; Holmes & Held, 2011)
  - For $j = 1, \ldots, J$, sample
  $$\frac{1}{\tau_{kj}} \sim \mathrm{iGauss}\left(\sqrt{\frac{\theta^2}{\beta_{kj}^2}}, \theta^2\right).$$

- Sample $\theta$ as follows:

$$\theta|\{\tau_{kj}\} \sim \mathrm{Gam}\left(Kp + c, \sum_k \sum_j \frac{\tau_{kj}}{2} + d\right). \quad (20)$$

The notation $\mathrm{iGauss}(a, b)$ denotes the inverse-Gaussian distribution with density given by

$$p(x) = \sqrt{\frac{b}{2\pi}} x^{-3/2} \exp\left(-\frac{b(x-a)^2}{2a^2 x}\right).$$